\documentclass[showpacs,amsmath,amssymb,aps,10pt,reprint,superscriptaddress,prb]{revtex4-1}
\usepackage{graphicx}
\usepackage{bm}
\usepackage[breaklinks=true,colorlinks=true,linkcolor=blue,urlcolor=blue,citecolor=blue]{hyperref}
\usepackage{graphics}
\usepackage{dcolumn}
\usepackage{amsmath,amssymb}
\usepackage{mathdots}
\usepackage{natbib}
\usepackage{soul,color}
\usepackage{epstopdf}
\usepackage[note-name]{notes2bib}

\begin{document}

\title{Annealing effects on the normal-state resistive properties of underdoped cuprates}

\author{R. V. Vovk}
\author{G. Ya. Khadzhai}
\author{Z. F. Nazyrov}
\author{S.\,N. Kamchatnaya}
    \affiliation{Physics Department, V. Karazin Kharkiv National University, 61077 Kharkiv, Ukraine}
\author{A. Feher}
    \affiliation{Pavol Josef \u{S}af\'{a}rik University, Park Angelinum 9, 04154 Ko\u{s}ice, Slovakia}
\author{O.~V.~Dobrovolskiy}
    \affiliation{Physikalisches Institut, Goethe University, 60438 Frankfurt am Main, Germany}
    \affiliation{Physics Department, V. Karazin Kharkiv National University, 61077 Kharkiv, Ukraine}

\begin{abstract}
The influence of room-temperature annealing on the parameters of the basal-plane electrical resistance of underdoped YBa$_2$Cu$_3$O$_{7-\delta}$ and HoBa$_2$Cu$_3$O$_{7-\delta}$ single crystals in the normal and superconducting states is investigated. The form of the derivatives $d\rho(T)/dT$ makes it possible to determine the onset temperature of the fluctuation conductivity and indicates a nonuniform distribution of the labile oxygen. Annealing has been revealed to lead to a monotonic decrease in the oxygen deficiency, that primarily manifests itself as a decrease of the residual resistance, an increase of $T_c$, and a decrease of the Debye temperature.
\end{abstract}
\maketitle

\section{Introduction}
Improving the stability of electronic transport characteristics is one of the most crucial applications-oriented problems in the contemporary physics of high-$T_c$ superconductivity. This problem is especially crucial for the most manufacturable and commonly used high-$T_c$ compounds from the so-called 1-2-3 system ReBa$_2$Cu$_3$O$_{7-\delta}$ (where $Re = Y$ or other rare earths) \cite{Wum87prl,Vov14ssc}.

\enlargethispage{2\baselineskip}
The presence of a labile component (oxygen) in these compounds often leads to a non-equilibrium state which can easily be induced by application of high hydrostatic pressure \cite{Sad00prb,Bal97ltp}, an abrupt temperature change \cite{Jor90pcs,Vov14ltp}, and a long-term aging \cite{Mar95apl,Lot10ltp}. These processes often lead to substantial structural changes \cite{Jor90pcs,Liz90ccs,Vov14jms1} of each particular sample, that, in turn, significantly affects the critical and electrophysical properties \cite{Kir93prb,Vov11jac,Gup95prb,Bon01ltp} of the system. In particular, the character of the temperature dependence of the conductivity may change from metal-like \cite{Bor91ssc,Vov13phb} to semiconductor-like \cite{Gin89boo,Vov14apa} one. This is accompanied by noticeable shifts of the temperature ranges of further peculiar phenomena such as the pseudogap anomaly \cite{Sad05prb,Vov15ssc}, fluctuation conductivity \cite{Fri89prb,Vov14cap}, metal-insulator transition \cite{Wid95pcs,Vov11jms}, non-coherent electronic transport \cite{And91prl,Vov09jms} and so on. According to current views \cite{Ash11snm,Vov09phb,Sol16prb} it is these non-trivial phenomena peculiar to the normal state which are expected to be the key to our understanding of the microscopic nature of high-$T_c$ superconductivity whose nature remains unclear despite an over-30-year-long history of extensive experimental and theoretical investigations \cite{Bed86zpb}.

Accordingly, in this work an analysis of the basal-plane conductivity in underdoped YBa$_2$Cu$_3$O$_{7-\delta}$ and HoBa$_2$Cu$_3$O$_{7-\delta}$ single crystals is performed in both, the normal and superconducting states. It is aimed at elucidation of the effect of annealing on the labile oxygen distribution, fluctuation conductivity, and charge carriers scattering in the normal state.

\section{Experimental}
The ReBa$_2$Cu$_3$O$_{7-\delta}$ single crystals (Re=Y, Ho) were grown by the solution-melt technique in a gold crucible as in Refs. \cite{Bal97ltp,Liz90ccs,Vov14jms1}. For resistive measurements three crystals K1, K2 (YBa$_2$Cu$_3$O$_{7-\delta}$) and K3 (HoBa$_2$Cu$_3$O$_{7-\delta}$) were selected. Electrical contacts were created in the standard 4-probe geometry by applying a silver paint on the crystal surface. This followed by attachment of silver conductors with $0.05$\,mm in diameter and a three-hour-long annealing at $200^\circ$C in ambient atmosphere. This procedure has allowed us to obtain a transient contact resistance of less than $1\,\Omega$ and to conduct resistance measurements at transport currents up to $10$\,mA in the $ab$-plane. The measurements were done in the temperature-sweep mode. Temperature was measured using a platinum resistor thermometer. The superconducting transition temperature was determined at the point of maxima in the dependences $d\rho_{ab}(T)/dT$ in the region of the superconducting transition.
\enlargethispage{2\baselineskip}

For reduction of the oxygen content the samples were annealed in an oxygen atmosphere at $680^\circ$C, $690^\circ$C (YBa$_2$ Cu$_3$O$_{7-\delta}$) and $600^\circ$C (HoBa$_2$Cu$_3$O$_{7-\delta}$) for 2-3 minutes. Once annealed the samples were quenched to room temperature within 2-3 minutes, mounted on the holder, and cooled down to liquid nitrogen temperatures within 10-15 minutes. All measurements were done while warming the samples up. For investigations of the effect of room-temperature annealing, after the first measurement of $\rho(T)$, the samples were kept at room temperature for several hours and the measurements were repeated. The final series of measurements was done after a room-temperature annealing of the samples for 5 days.

\section{Results and Discussion}
\subsection{Influence of defects on the fluctuation conductivity and phonon scattering}

In our works \cite{Vov14phb,Vov15pcs} it has been shown that the temperature dependence of the in-plane normal-state resistivity of the high-$T_c$ 1-2-3 system, $\rho_n(T)$ can be approximated well by the Bloch-Gr\"uneisen formula describing the charge carriers scattering on phonons and defects
\begin{equation}
\label{eBG}
    \rho_n(T) = \rho_0 + C_3\left(\frac{T}{\theta}\right)^3 \int_{0}^{\theta/T}\frac{e^x x^3 dx}{(e^x - 1)^2}.
\end{equation}
Here $\rho_0$ is the residual resistivity due to the defects, $C_n$ is the phonon scattering coefficient, $\theta$ is the Debye temperature, and $n =3$ that corresponds \cite{Col65jap} to interband scattering.
\begin{figure}[t!]
    \centering
    \includegraphics[width=0.9\linewidth]{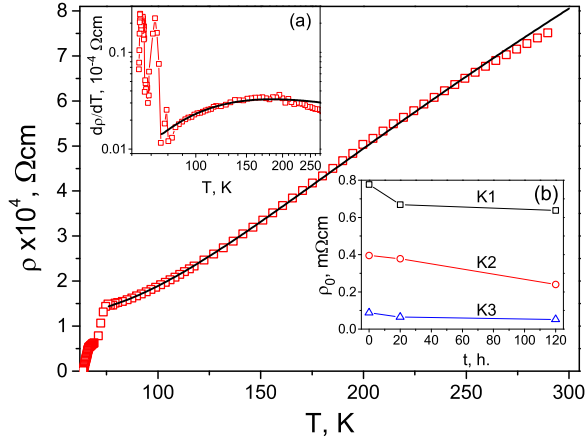}
    \caption[]
    {Temperature dependence of the basal-plane resistance of sample K3 (HoBa$_2$Cu$_3$O$_{7-\delta}$) after quenching from $600^\circ$C. The symbols denote the experimental data while the solid line is a fit of the normal-state resistance to Eq. (\ref{eBG}). Inset (a): Temperature dependence of the derivative $d\rho_{ab}(T)/dT$ for the same sample in the whole investigated temperature range. The solid line is calculated by Eq. (\ref{eBG}). Inset (b): Dependences of the residual resistivity $\rho_0$ on the annealing time for all investigated samples. Solid lines are guides for the eye.}
    \label{f1}
\end{figure}

Equation (\ref{eBG}) describes the experimental curve $\rho_{ab}(T)$ very well, refer to Fig. \ref{f1}. The average error in the temperature range from $T_c$ to $300$\,K does not exceed $1\%$. The derivative $d\rho_{ab}(T)/dT$ calculated by Eq. (\ref{eBG}) qualitatively agrees with the derivative calculated from the experimental dependence $\rho_{ab}(T)$. The systematic deviations may be associated with inhomogeneity of the sample and the fact that Eq. (\ref{eBG}) does not account for all mechanisms of charge carriers scattering \cite{Apa02prb65,Vov03prb,Ada94ltp,Vov03prl,Cur11prb}.

In particular, at high temperatures the experimental values $\rho_{ab}(T)$ and $d\rho_{ab}(T)/dT$ turn down from the approximation by Eq. (\ref{eBG}) that points to a tendency to saturation. This tendency is typical for many transition metals and their alloys, including superconducting ones. The mechanisms of resistance saturation are discussed, e.g. in Refs. \cite{Ais70psj,All80boo,Cla72thp,Zhu89boo,Gan13boo,San84pra}.

The approximation parameters for Eq. (\ref{eBG}) for all $\rho_{ab}(T)$ analyzed in this work are reported in Table \ref{table}.
\begin{table}[tbh!]
\centering
{\footnotesize
   \begin{tabular}{|c|c|c|c|}
\hline
                & Quenching                 & Annealing         & Annealing \\
                & from $600^\circ$C (a)   & at $20^\circ$C for      & at $20^\circ$C for\\
                 &                        & 20 hours (b)    & 120 hours (c)\\
\hline
\multicolumn{4}{|c|}{K1 YBa$_2$Cu$_3$O$_{7-\delta}$}\\
\hline
$T_{c1}$,\,K     & 36.7(0) & 39.975(8.9) & 40.675(10.8)\\
$T_{c2}$,\,K     & 43.7(0) & 43.4(-0.7)  & 43.8(0.2)\\
$T_{c3}$,\,K     & 49(0)   & 46.55(-5.0) & 47.05(-4.0)\\
$T_{c4}$,\,K     & ---     & 50.4(-5.0)  & 49.35(-2.1)\\
$\Delta T_{0.5-1}$,\,K     & 3.525(0)    & 2.20(-37.6) & 2.20(-37.6)\\
$\Delta T_{0.5-2}$,\,K     & 2.94(0)     & 1.531(-47.9)  & 1.50(-49.0)\\
$\Delta T_{0.5-3}$,\,K     & 3.20(0)     & 2.01(-37.2) & 1.33(-58.4)\\
$\Delta T_{0.5-4}$,\,K     & ---        & 3.525(0)  & 6.41(81.8)\\
$\rho_0$,\,m$\Omega$cm     & 0.7765(0)     & 0.669(-13.8)  & 0.637(-18.0)\\
$C_3$,\,m$\Omega$cm        & 10.3(0)     & 9.495(-7.8) & 9.315(-9.6)\\
$\theta$,\,K               & 913.5(0)    & 912(-0.2)  & 924.5(1.21)\\

\hline
\multicolumn{4}{|c|}{K2 YBa$_2$Cu$_3$O$_{7-\delta}$}\\
\hline
$T_{c1}$,\,K             & 43.950(0) & 47.7(8.5) & 48.45(10.2)\\
$T_{c2}$,\,K             & 49.35(0)  & 51.4(4.2) & 52.0(5.4)\\
$\Delta T_{0.5-1}$,\,K   & 1.602(0)  & 2.275(42.0) & 2.137(33.4)\\
$\Delta T_{0.5-2}$,\,K   & 3.917(0)  & 2.352(40.0) & 3.525(35.5)\\
$\rho_0$,\,m$\Omega$cm   & 0.395(0)  & 0.378(-4.3) & 0.2395(-39.4)\\
$C_3$,\,m$\Omega$cm      & 8.87(0)   & 8.995(1.4)  & 8.425(5.0)\\
$\theta$,\,K             & 910(0)    & 957.5(5.2)  & 970(6.6)\\

\hline
\multicolumn{4}{|c|}{K3 HoBa$_2$Cu$_3$O$_{7-\delta}$}\\
\hline
$T_{c1}$,\,K             & 64.2(0) & 70.08(9.2) & 73.91(15.1)\\
$T_{c2}$,\,K             & 65.6(0) & 72.10(0)   & --- \\
$T_{c3}$,\,K             & 71.9(0) & 74.80(0)   & 77.90(0)\\
$\Delta T_{0.5-1}$,\,K   & 0.95(0) & 1.24(0)    & 1.98(0)\\
$\Delta T_{0.5-2}$,\,K   & 1.41(0) & 1.66(0)    & ---\\
$\Delta T_{0.5-3}$,\,K   & 4.15(0) & 2.94(0)    & 2.33(0)\\
$\rho_0$,\,m$\Omega$cm     & 0.088(0)     & 0.06425(-27.0)  & 0.0522(-40.7)\\
$C_3$,\,m$\Omega$cm        & 3.19(0)     & 2.5(0) & 1.95(0)\\
$\theta$,\,K               & 576.5(0)    & 620.5(0)  & 583(0)\\
\hline
\end{tabular}
}
   \caption{Heat processing, the approximation parameters for $\rho_{ab}(T)$ in the normal state by Eq. (\ref{eBG}), and the characteristics of the superconducting transition by Eq. (\ref{eDer}).}
   \label{table}
\end{table}

According to Eq. (\ref{eBG}) the derivative $d\rho_{n}/dT$ exhibits a smeared maximum at $T_{max} = 0.35\theta$ that corresponds to $T_{max} \approx 100\div 200$\,K. Below $T_{max}$ the resistivity temperature derivative exhibits a sharp maximum at $T_c$ as the sample transits into the superconducting state, $d\rho_{sc}/dT|_{T_c} \gg d\rho_{n}/dT|_{T_{max}}$. Between these two maxima one sees a minimum in $d\rho/dT$ which points to the onset of the superconducting transition, that is to a transition to the regime of fluctuation conductivity. It should be noted that the minimum in $d\rho/dT$ is seen very well for underdoped samples where $T_{max}$ essentially exceeds $T_c$. For optimally doped samples with $T_{max}\geq T_c$ this minimum is not necessarily observed. In this case, the onset temperature for the fluctuation conductivity can be estimated from the data of Refs. \cite{Ler01prl,Vov17phb}.

The symbols in Fig. \ref{f1} depict the dependence $\rho(T)$ for one of the samples (K3 after quenching from $600^\circ$C), while the solid curve is a fit of $\rho(T)$ to Eq. \ref{eBG}. The derivative $d\rho/dT$ for the same sample is plotted in inset (a) in Fig. \ref{f1}. One recognizes a maximum in $d\rho/dT$ at about $165$\,K, a minimum in $d\rho/dT$ at about $80$\,K, below which the regime of fluctuation conductivity sets on, and maxima in the vicinity of the superconducting transition.

In inset (b) of Fig. \ref{f1} the dependences of the residual resistivity $\rho_0$ on the annealing time are displayed. Since $\rho_0$ is associated with defects, one can assume that these defects are mostly oxygen vacancies. The reduction of $\rho_0$ attests to a reduction of the number of defects in the course of annealing.

In Table \ref{table} one sees that $\rho_0$ changes for the investigated samples in different ranges that allows one to plot dependences of the approximation parameters according to Eq. (\ref{eBG}) as well as the onset temperature of the fluctuation conductivity $T_{fluct}$ on the residual resistivity.

\begin{figure}
    \centering
    \includegraphics[width=0.9\linewidth]{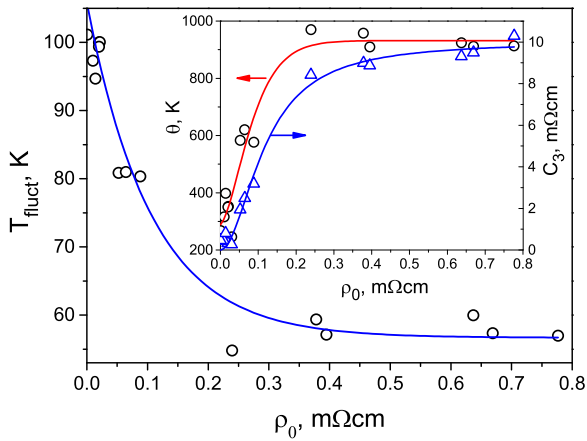}
    \caption[]
    {Fluctuation conductivity onset temperature $T_{fluct}$ versus residual resistivity $\rho_0$. Inset: Dependence of the Debye temperature $\theta$ and the phonon scattering coefficient $C_3$ (both as approximation parameters) on $\rho_0$. The data are presented for the HoBa$_2$Cu$_3$O$_{7-\delta}$ crystal with $0.05 < \rho_0 < 0.09$\,(m$\Omega$cm) investigated in this work and two  YBa$_2$Cu$_3$O$_{7-\delta}$ single crystals with $0.2 < \rho_0 < 0.4$\,(m$\Omega$cm) and $0.6 < \rho_0 < 0.8$\,(m$\Omega$cm) from Ref. \cite{Vov14phb,Vov15pcs}. Lines are guides for the eye.}
    \label{f2}
\end{figure}

The resulting dependences are displayed in Fig. \ref{f2}. One should note that the data for HoBa$_2$Cu$_3$O$_{7-\delta}$ fall onto the data for YBa$_2$Cu$_3$O$_{7-\delta}$, that is substitution of Y for Ho does not substantially affect the electronic transport properties of the considered system not only in the case of optimally-doped samples with $\delta \leq0.15$ \cite{Kha14fnt}, but also in the case of underdoped samples. The latter finding agrees with the experimental data of Ref. \cite{Won06jrn} where it was observed that in compounds of the 1-2-3 system, substitution of Y by lanthanides of smaller ion radii such as Ho and Er affects the dependence $T(\delta)$ only slightly.

In Fig. \ref{f2} one sees that the reduction of $\rho_0$ is associated with an increase of $T_{fluct}$ up to $T_{fluct} \approx 100$\,K, that is with a transition of the system into the regime of optimal doping \cite{Ler01prl,Vov17phb}. This supports the assumption that $\rho_0$ is primarily associated with oxygen vacancies.

In the inset to Fig. \ref{f2} one sees that the phonon scattering coefficient $C_3$ decreases with decreasing $\rho_0$ that is with improvement of the lattice quality. A similar effect was previously observed, e.g. in Ref. \cite{Kho83fnt}. For the system investigated by us the change of the phonon scattering coefficient can be associated with the deformation of the phonon spectrum of the sample in the presence of defects, see e.g. \cite{Kag71etp}, which are represented by oxygen vacancies in our case.

The changes of the Debye temperatures, which also decreases with decreasing $\rho_0$, are in line with the discussion above. Since $\Delta \theta / \theta \approx - \alpha \Delta V / V + \beta \delta f / f$ ($\Delta V$ is the change of the volume of the unit cell and $\Delta f$ is the change of the force constants), the primary changes of $\theta$ at the crossover from the optimal doping to a large oxygen deficiency are largely stipulated by the increase of the force constants, that is associated with the deformation of the phonon spectrum.

\subsection{Inhomogeneity of the superconducting state in underdoped samples of the 1-2-3 system}

As is known \cite{Vov17phb,Kha17cap} for underdoped samples from the 1-2-3 system, a series of narrow and high peaks is observed in $d\rho/dT$ below $T_c$. This points to a subsequential transition in the superconducting state of regions with the different $T_c$s, that is different $\delta$. The resistance vanish attests to that at some temperature a superconducting region appears, which spreads over the entire sample volume and shunts all other superconducting regions with lower $T_c$s and/or non-superconducting regions. As is well known, these regions may not be considered as connected in series or in parallel in the general case \cite{Ros03boo}. Therefore, the experimental curves $\rho(T)$ characterize some effective resistivity values.

Maxima in $d\rho/dT$ in the vicinity of the superconducting transition can be described as \cite{Rol83boo}
\begin{equation}
\label{eDer}
    \frac{d\rho_i}{dT}\approx \rho_{in}\displaystyle\frac{\displaystyle\left(\frac{\displaystyle\frac{1}{w_i}\exp(T_{ci} - T)}{w_i}\right)}{\left(1 + \displaystyle\frac{\exp(T_{ci} - T)}{w_i}\right)^2},
\end{equation}
where $T_{ci}$ and $w_i$ characterize, respectively, the temperature of the superconducting transition in the $i$-th region and its width (the width of the maximum in $d\rho/dT$ at half height amounts to $\Delta T_{c_{0.5i}} \approx 3.5w_i$). Further, $\rho_i \approx const$ is the normal-state resistivity of the $i$-th region just above the superconducting transition. The parameters of the maxima in $d\rho_i(T)/dT$ according to Eq. (\ref{eDer}) for the investigate samples in the superconducting transition region are reported in Table \ref{table}.
\begin{figure}
    \centering
    \includegraphics[width=0.95\linewidth]{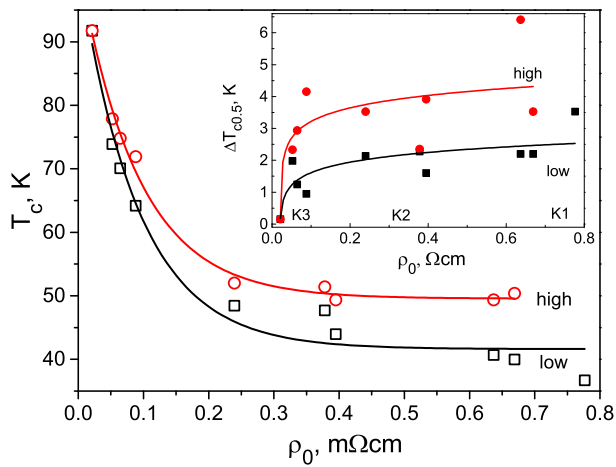}
    \caption[]
    {Correlations of the most high $T_{c\,high}$ and low $T_{c\,low}$ from the superconducting transition temperatures with the residual resistivity $\rho_0$ for the HoBa$_2$Cu$_3$O$_{7-\delta}$ crystal with $0.05 < \rho_0 < 0.09$\,(m$\Omega$cm) investigated in this work and two  YBa$_2$Cu$_3$O$_{7-\delta}$ single crystals with $0.2 < \rho_0 < 0.4$\,(m$\Omega$cm) and $0.6 < \rho_0 < 0.8$\,(m$\Omega$cm) from Refs. \cite{Vov14phb,Vov15pcs}. Inset: The same for the respective superconducting transition widths, $\Delta T_{c0.5i}$. Lines are guides for the eye.}
    \label{f3}
\end{figure}

The number of these maxima in $d\rho(T)/dT$ for the different samples varies and it is not conserved in the course of aging. However, in all cases there exist the most high- and low-temperature maxima, whose parameters correlate with the residual resistivity $\rho_0$. These correlations are displayed in Fig. \ref{f3}. As for the normal state, here the data for HoBa$_2$Cu$_3$O$_{7-\delta}$ and YBa$_2$Cu$_3$O$_{7-\delta}$ are lying on the same curves, that is substitution of Y by Ho does not noticeably affect the characteristics of the superconducting transition.
\begin{figure}
    \centering
    \includegraphics[width=0.95\linewidth]{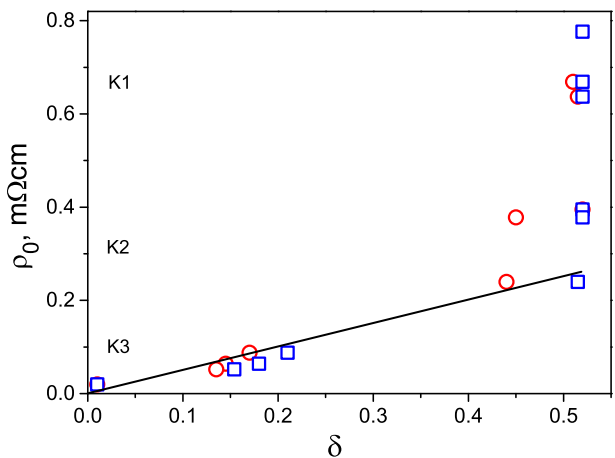}
    \caption[]
    {Correlation $\rho_0$ -- $\delta$ for the samples studied, constructed on the basis of the dependences $T_c(\delta)$ from Ref. \cite{Won06jrn}.
    $\square - T_{c~low}$, $\bigcirc - T_{c~high}$.}
    \label{f4}
\end{figure}

One sees that with decreasing $\rho_0$, that is with a decrease of the degree of disorder of the sample, both $T_{ci}$ increase up to the values typical for optimally doped samples. This means that in the course of annealing the oxygen content increases and tends to its optimal value. We believe, this process is stipulated by the coalescence of clusters of oxygen vacancies in the sample, which in the very end should disappear after going out onto the crystal surface \cite{Kha17cap}. The coalescence of clusters of nonstochiometric oxygen vacancies leads to a narrowing of the maxima in $d\rho(T)/dT$: The widths $\Delta T_{c0.5i}$ decrease with decreasing $\rho_0$, refer to the inset in Fig. \ref{f3}.

The correlation between $T_c$ and $\rho_0$ allows us to present the respective correlation between $\rho_0$ and $\delta$. In Fig. \ref{f4} this correlation is plotted on the basis of the data of Ref. \cite{Won06jrn}. One sees that for $\delta < 0.5$ the residual resistivity is proportional to the oxygen index $\rho_0 \approx a\delta$. This means that the residual resistivity caused by nonstoichiometric vacancies is noticeably larger than that due to other defects. At $\delta \geq 0.5$ one recognizes a nearly vertical drop of $T_c$ in the dependence $T_c(\delta)$, which is why it is impossible to extend the correlation between $T_c$ and $\rho_0$ into this range. Probably, at such values of $\delta$, $\rho_0$ depends not only on the defect concentration but also on the density of the charge carriers, which here changes.

\section{Conclusion}
The conducted analysis of the temperature dependence of the basal-plane resistance of HoBa$_2$Cu$_3$O$_{7-\delta}$ and  YBa$_2$Cu$_3$O$_{7-\delta}$ single crystals in the normal and superconducting states allows us to conclude the following:
The behavior of $d\rho(T)/dT$ in the normal state attests to the charge scattering on phonos and defects.
In the superconducting state, the presence of several derivative maxima $d\rho(T)/dT$ indicates an inhomogeneous oxygen distribution.
The sharp minimum in $d\rho(T)/dT$ points to the onset of the fluctuation conductivity.
The scattering parameters in the normal state, the fluctuation conductivity range and the characteristics of the superconducting state correlate with the residual resistivity $\rho_0$.
For $\delta < 0.5$ the residual resistivity is determined by the oxygen index.

\section*{Acknowledgements}
The research leading to these results has received funding from the European Union's Horizon 2020 research and innovation program under Marie Sklodowska-Curie Grant Agreement No. 644348 (MagIC).


\end{document}